\documentclass[prc,floatfix,groupedaddress,nofootinbib,showpacs,preprintnumbers,
amsmath,amssymb,amsfonts,superscriptaddress,widetable] {revtex4}
\usepackage{graphicx}% Include figure files
\usepackage{dcolumn}%Align table columns on decimal point
\usepackage{mathrsfs}
\usepackage{bm}% bold math

\usepackage{graphicx}% Include figure files
\usepackage{dcolumn}% Align table columns on decimal point
\usepackage{bm}% bold math
\usepackage[usenames]{color}
\begin{document}

\title{Accurate calibration of relativistic mean-field models:\\
correlating observables and providing meaningful theoretical
uncertainties}
\author{F. J. Fattoyev}
\email{ff07@fsu.edu} \affiliation{Department of Physics, Florida
State University, Tallahassee, FL 32306} \affiliation{Institute of
Nuclear Physics, Tashkent 100214, Uzbekistan}
\author{J. Piekarewicz}
\email{jpiekarewicz@fsu.edu}
\affiliation{Department of Physics, Florida State University,
Tallahassee, FL 32306}

\date{\today}
\begin{abstract}
 Theoretical uncertainties in the predictions of relativistic
 mean-field models are estimated using a chi-square minimization
 procedure that is implemented by studying the small oscillations
 around the chi-square minimum.  By diagonalizing the matrix of second
 derivatives, one gains access to a wealth of information---in the form
 of powerful correlations---that would normally remain hidden. We
 illustrate the power of the {\sl covariance analysis} by using two relativistic
 mean-field models: (a) the original linear Walecka model and (b) the
 accurately calibrated FSUGold parametrization. In addition to
 providing meaningful theoretical uncertainties for both model
 parameters and predicted observables, the covariance analysis
 establishes robust correlations between physical observables. In
 particular, we show that whereas the correlation coefficient
 between the slope of the symmetry energy and the neutron-skin
 thickness of Lead is indeed very large, a 1\% measurement of the
 neutron radius of Lead may only be able to constrain the slope of
 the symmetry energy to about 30\%.
\end{abstract}
 \maketitle

\section{Introduction}
\label{intro}
The need to provide meaningful uncertainties in theoretical
predictions of physical observables is a theme that is gaining
significant momentum among the scientific community. Indeed, the
search for a microscopic theory that both predicts and provides
well-quantified theoretical uncertainties is one the founding pillars
of the successful UNEDF collaboration~\cite{UNEDF}.  Moreover, as
recently articulated in an editorial published in the Physical Review
A~\cite{PhysRevA.83.040001}, theoretical predictions submitted for
publication are now expected to be accompanied by meaningful
uncertainty estimates. The need for {\sl ``theoretical error bars''}
becomes particularly critical whenever models calibrated in certain
domain are used to extrapolate into uncharted regions.

Although firmly rooted in QCD, computing both the nucleon-nucleon
(NN) interaction and the properties of nuclei in terms of the underlying
quark and gluon constituents remains a daunting task. Hence, rather
than relying strictly on QCD, one uses the properties of QCD (such as
chiral symmetry and relevant energy scales) as a guide to construct
phenomenological interactions using nucleons and mesons as the
fundamental degrees of freedom. However, QCD has little to say about
the strength of the underlying model parameters which must then be
constrained from experimental data. For example, deuteron properties
along with two-body scattering data are used to build a
nucleon-nucleon interaction that may then be used (supplemented
with a phenomenological three-body force) to compute {\sl ab-initio}
the properties of light nuclei.  Attempting ab-initio calculations of
the properties of medium-to-heavy nuclei remains well beyond the scope
of the most powerful computers to date.  In this case one must bring to
bear the full power of density functional theory (DFT).  Following the
seminal work by Kohn and collaborators~\cite{Kohn:1999}, DFT shifts
the focus from the complicated many-body wave-function to the much
simple one-body density. Moreover, Kohn and Sham have shown how the
one-body density may be obtained from a variational problem that
reduces to the solution of a set of mean-field-like ({\sl
``Kohn-Sham''}) equations~\cite{Kohn:1965}.
The form of the Kohn-Sham potential is in
general reminiscent of the underlying (bare) NN potential.  However,
the constants that parametrize the Kohn-Sham potential are directly
fitted to many-body observables (such as masses and charge radii)
rather than two-body data.  In this manner the complicated dynamics
originating from exchange and correlation effects get implicitly
encoded in the empirical constants.  Yet regardless of whether the
effective interaction is fitted to two-nucleon or to many-body
data---the determination of the model parameters often relies on the
optimization of a quality measure.

In this contribution we focus on density functional theory and follow
the standard protocol of determining the model parameters through a
$\chi^{2}$-minimization procedure.  This procedure is implemented by:
(a) selecting a set of accurately measured ground-state observables
and (b) demanding that the differences between these observables and
the predictions of the model be minimized. Note that in the present
framework a model consists of both a set of parameters and a
$\chi^{2}$-measure. In general, modifying the $\chi^{2}$-measure
({\sl e.g.,} by adding observables) results in a change in the model
parameters. Traditionally, once the $\chi^{2}$-minimum has been found
one proceeds to validate the model against observables not included in
the quality fit. Nuclear collective excitations are a potentially
{\sl ``safe''} testing arena for the model as they represent the
{\sl small oscillations} around the variational ground state.
But what happens when the model must be extrapolated to regions
of large isospin imbalance and high density as in the interior of
neutron stars? Clearly, without reliable theoretical uncertainties it
is difficult to assess the predictions of the model. To remedy this
situation we propose to study the {\sl small oscillations} around the
$\chi^{2}$-minimum---rather than the minimum itself.  As we shall see,
such a statistical analysis---inspired by the recent study reported in
Ref.~\cite{Reinhard:2010wz}---provides access to a wealth of
information that remains hidden if one gets trapped in the
$\chi^{2}$-minimum.  Among the critical questions that we will be able
to answer is how fast does the $\chi^{2}$-measure deteriorate as one
moves away from the minimum.  Should additional observables be added
to the $\chi^{2}$-measure to better constrain the model? And if such
observables are hard to determine are there others that may be more
readily accessible and provide similar constraints? A particularly
topical example that illustrates such a synergy is the correlation
between the neutron-skin thickness and electric dipole polarizability
of neutron-rich nuclei~\cite{Reinhard:2010wz,Piekarewicz:2010fa}. A
detailed analysis of such correlation---which involves a systematic
study of the isovector dipole response---is beyond the scope of this
initial study and will become the subject of a forthcoming
publication. Yet the study of correlations among observables sensitive
to the poorly-determined density dependence of the symmetry energy
will become a recurring theme throughout this contribution.

The manuscript has been organized as follows. In Sec.~\ref{Formalism}
we develop the necessary formalism to implement the correlation
analysis. This section is divided in two parts: (a) a discussion on
the structure of a class of relativistic mean-field models and
(b) a relatively short---yet fairly complete---derivation of the
statistical formalism required to perform the covariance analysis.
In Sec.~\ref{Results} two simple examples are used to illustrate the
power of the formalism. This exercise  culminates with the estimation
of meaningful theoretical error bars and correlation coefficients.
Our conclusions and outlook are presented in Sec.~\ref{Conclusions}.

\section{Formalism}
\label{Formalism}

In this section we develop the formalism required to implement the
correlation analysis. First, in Sec.~\ref{RMF} we introduce a fairly
general class of relativistic mean-field models that are rooted in
effective-field-theory concepts, such as naturalness and power
counting. Second, in Sec.~\ref{LRCA} we present a self-contained
derivation of the ideas and formulas required to implement the
covariance analysis.

\subsection{Relativistic Mean-Field Models}
\label{RMF}

Relativistic mean-field models traditionally include an isodoublet
nucleon field ($\psi$) interacting via the exchange of two
isoscalar mesons (a scalar $\phi$ and a vector $V^{\mu}$), one
vector-isovector meson ($b^{\mu}$), and the photon
($A^{\mu}$)~\cite{Serot:1984ey, Serot:1997xg,Mueller:1996pm}. The
non-interacting Lagrangian density for such a model may be written
as follows:
\begin{align}
{\mathscr L}_{0} & =
\bar\psi \left(i\gamma^{\mu}\partial_{\mu}\!-\!M\right)\psi +
\frac{1}{2}\partial_{\mu}\phi\,\partial^{\mu} \phi
-\frac{1}{2}m_{\rm s}^{2}\phi^{2} \nonumber \\
&-\frac{1}{4}V^{\mu\nu}V_{\mu\nu} + \frac{1}{2}m_{\rm
v}^{2}V^{\mu}V_{\mu} - \frac{1}{4}{\bf b}^{\mu\nu}\cdot{\bf
b}_{\mu\nu} + \frac{1}{2}m_{\rho}^{2}\,{\bf b}^{\mu}\cdot{\bf
b}_{\mu} -\frac{1}{4}F^{\mu\nu}F_{\mu\nu} \;, \label{Lagrangian0}
\end{align}
where $V_{\mu\nu}$, ${\bf b}_{\mu\nu}$, and $F_{\mu\nu}$
are the isoscalar, isovector, and electromagnetic field
tensors, respectively. That is,
\begin{subequations}
\begin{align}
 V_{\mu\nu} &= \partial_{\mu}V_{\nu} - \partial_{\nu}V_{\mu} \;, \\
 {\bf b}_{\mu\nu} &= \partial_{\mu}{\bf b}_{\nu}
 - \partial_{\nu}{\bf b}_{\mu} \;, \\
F_{\mu\nu} &= \partial_{\mu}A_{\nu} - \partial_{\nu}A_{\mu} \;.
\label{FieldTensors}
\end{align}
\end{subequations}
The four constants $M$, $m_{\rm s}$, $m_{\rm v}$, and $m_{\rho}$
represent the nucleon and meson masses and may be treated (if
wished) as empirical parameters. Often, however, $m_{\rm s}$ is
determined from an accurate calibration procedure.  The interacting
Lagrangian density has evolved significantly over the years and now
incorporates a variety of meson self-interacting terms that are
designed to improve the quality of the model. Following ideas
developed in Ref.~\cite{Mueller:1996pm} we write the interacting
Lagrangian density in the following form:
%%%
\begin{equation}
{\mathscr L}_{\rm int} =
\bar\psi \left[g_{\rm s}\phi   \!-\!
         \left(g_{\rm v}V_\mu  \!+\!
    \frac{g_{\rho}}{2}\mbox{\boldmath$\tau$}\cdot{\bf b}_{\mu}
                               \!+\!
    \frac{e}{2}(1\!+\!\tau_{3})A_{\mu}\right)\gamma^{\mu}
         \right]\psi -
          U(\phi,V_{\mu},{\bf b_{\mu}}) \;.
\label{Lagrangian}
\end{equation}
%%%
In addition to the standard Yukawa interactions, the Lagrangian is
supplemented with an effective potential
$U(\phi,V_{\mu},{\bf b_{\mu}})$ consisting of non-linear meson
interactions that serve to simulate the complicated dynamics that lies
beyond the realm of the mean-field theory. Indeed, by fitting the
various coupling constants directly to nuclear properties---rather
than to two-nucleon data---the complicated dynamics originating from
nucleon exchange, short-range effects, and many-body correlations gets
implicitly encoded in a small number of parameters.  For the purpose
of the present discussion we introduce explicitly all non-linear terms
up to fourth-order in the meson fields. That is,
%%%
\begin{align}
  U(\phi,V^{\mu},{\bf b}^{\mu})  &=
    \frac{\kappa}{3!} \Phi^3
     \!+\!\frac{\lambda}{4!}{\Phi}^4
     \!-\!\frac{\zeta}{4!}\Big(W_{\mu}W^\mu\Big)^2
     \!-\!\Lambda_{\rm v}\Big(W_{\nu}W^\nu\Big)
       \Big({\bf B}_{\mu}\cdot{\bf B}^{\mu}\Big)
     \!-\!\frac{\xi}{4!}\Big({\bf B}_{\mu}\cdot{\bf B}^\mu\Big)^2
      \\ \nonumber &
     \!+\!\kappa_{0}\Phi W_{\mu}W^{\mu}
     \!+\!\kappa_{1}\Phi{\bf B}_{\mu}\cdot{\bf B}^\mu
     \!+\!\lambda_{0}\Phi^2W_{\mu}W^{\mu}
     \!+\!\lambda_{1}\Phi^2{\bf B}_{\mu}\cdot{\bf B}^{\mu}
     \!-\!\Lambda^{\prime}_{\rm v}\Big(W_{\mu}W_{\nu}\Big)
       \Big({\bf B}^{\mu}\cdot{\bf B}^{\nu}\Big)
    \!+\!\ldots
\label{USelf}
\end{align}
%%%
where the following definitions have been introduced:
$\Phi\!\equiv\!g_{\rm s}\phi$,
$W_{\mu}\!\equiv\!g_{\rm v}V_{\mu}$, and
${\bf B}_{\mu}\!\equiv\!g_{\rho}{\bf b}_{\mu}$. Given that
the present analysis will be restricted to the study of
uniform nuclear matter, terms proportional to the
derivatives of the meson fields have not been included.
As it stands, the relativistic model contains 14
undetermined parameters (1 meson mass, 3 Yukawa
couplings, and 10 meson self-interaction terms).  Note
that if one incorporates the occasionally-used scalar-isovector
$\delta$-meson~\cite{Liu:2001iz,Baran:2004ih}, then 9
additional parameters must be included to this order (1
Yukawa coupling, and 8 meson self-interaction terms).

A model with 14---or 23---parameters goes significantly beyond
the early relativistic models that were able to reproduce the
saturation point of symmetric nuclear matter as well as various
ground-state observables with only a handful of parameters (a single
meson mass and three Yukawa couplings)~\cite{Walecka:1974qa,
Horowitz:1981xw,Serot:1984ey}. Although fairly successful, those
early models suffered from a major drawback: an unrealistically
large incompressibility coefficient. Such a problem was successfully
solved by Boguta and Bodmer with the introduction of cubic and
quartic scalar meson self-interactions~\cite{Boguta:1977xi}.
Remarkably, using only these six parameters ($m_{\rm s}, g_{\rm s},
g_{\rm v}, g_{\rho}, \kappa,\lambda$) it is possible to reproduce a
host of ground-state properties of finite nuclei (both spherical and
deformed) throughout the periodic
table~\cite{Lalazissis:1996rd,Lalazissis:1999}. And by adding two
additional parameters ($\zeta$ and $\Lambda_{\rm v}$) the
success of the model can be extended to the realm of nuclear
collective excitations and neutron-star
properties~\cite{Horowitz:2000xj,Todd-Rutel:2005fa,Piekarewicz:2007dx,
Fattoyev:2010mx}.

Given that the existent database of both laboratory and observational
data appears to be accurately described by an 8-parameter model, is
there any compelling reason to include 6---or 15---additional
parameters?  And if so, what criteria does one use to constrain these
remaining parameters?  A meaningful criterion used to construct
an effective Lagrangian for nuclear-physics calculations has been
proposed by Furnstahl, Serot, and Tang based on the concept of
{\sl ``naive dimensional analysis''} and
{\sl ``naturalness''}~\cite{Furnstahl:1996wv, Furnstahl:1996zm}.
The basic idea behind naturalness is that once the {\sl dimensionful}
meson fields (having units of mass) have been properly scaled using
strong-interaction mass scales, the remaining {\sl dimensionless}
coefficients of the effective Lagrangian should all be
{\sl ``natural''}; that is, neither too small nor too
large~\cite{Furnstahl:1996wv,Agrawal:2010wg}.  Such an approach is
both useful and powerful as it allows an organizational scheme based
on an expansion in powers of the meson fields. Terms
in the effective Lagrangian with a large number of meson fields will
then be suppressed by a large strong-interaction mass scale. In this
regard the assumption of naturalness is essential as the suppression
from the large mass scale should not be compensated by large, {\sl i.e.,}
unnatural, coefficients. It was by invoking the concept of naturalness
that we were able to truncate the effective potential
$U(\phi,V^{\mu},{\bf b}^{\mu})$  beyond quartic terms in the meson
fields.

Although we have justified the truncation of the effective Lagrangian
invoking naturalness, we are not aware of an additional organizational
principle that may be used {\sl a-priori} to limit further the form of
$U(\phi,V^{\mu},{\bf b}^{\mu})$. This implies that {\sl all} model
parameters must be retained, as it is unnatural to set some
coefficients arbitrarily to zero without a compelling symmetry
argument~\cite{Furnstahl:1995zb}.  In principle then, all model
parameters must be retained and subsequently determined from a
fit to empirical data. In practice, however, many successful theoretical
models---such as NL3~\cite{Lalazissis:1996rd,Lalazissis:1999}
and FSUGold~\cite{Todd-Rutel:2005fa}---arbitrarily set some
of these parameters to zero.  The {\sl ``justification''} behind these
fairly ad-hoc procedure is that whereas the neglected terms are of the
same order in a power-counting scheme, the full set of parameters is
poorly determined by existing data, so ignoring a subset model
parameters does not compromise the quality of the
fit~\cite{Mueller:1996pm,Furnstahl:1996wv}.

An important goal of the present work is to investigate correlations
among the parameters of the model and whether additional physical
observables could remove such correlations.
To do so we follow the standard protocol of
determining the model parameters through a $\chi^{2}$ minimization
procedure. Traditionally, this procedure is implemented by selecting
a set of accurately measured ground-state observables for a variety of
nuclei and then demanding that the differences between the observables
and the predictions of the model be minimized. Once this is done, the
success of the model may be gauged by computing observables not
included in the fit. However, it is often difficult to assess the
uncertainty in the predictions of the model. To address this
deficiency we propose to study the {\sl small oscillations} around the
minimum---rather than the minimum itself.  Such a
study---inspired by the recent statistical analysis presented in
Ref.~\cite{Reinhard:2010wz}---provides access to a wealth of
information that, in turn, enables one to specify meaningful
theoretical error bars as well as to explore correlations among
model parameters and calculated observables.

Although the following discussion is framed in the context of an
underlying $\chi^{2}$-measure, our arguments are general as they
merely rely on the existence of a (local) minimum (or an extremum).
As in any small-oscillations problem, deviations of the
$\chi^{2}$-measure from its minimum value are controlled by a symmetric
$F\!\times\!F$ matrix, where $F$ represents the total number of model
parameters. Being symmetric, such a matrix may be brought into a
diagonal form by means of an orthogonal transformation. The outcome of
such a diagonalization procedure is a set of $F$ eigenvalues and $F$
eigenvectors. When a point in parameter space is expanded in terms of
these eigenvectors, the deviations of the $\chi^{2}$-measure from its
minimum value take the form of a system of $F$ {\sl uncoupled harmonic
oscillators}---with the eigenvalues playing the role of the $F$ spring
constants. The spring constants may be {\sl ``stiff''} or {\sl
``soft''} depending on whether the curvature around the minimum is
steep or shallow, respectively. As one explores the parameter
landscape along a stiff direction---and thus along a particular linear
combination of model parameters---a rapid worsening of the
$\chi^{2}$-measure ensues, suggesting that the fitting protocol is
robust enough to constrain this particular linear combination.
Conversely, no significant deterioration in the quality of the fit is
observed as one moves along a soft direction. In this case the
$\chi^{2}$-minimum is of little significance as scores of parameter
sets ({\sl i.e.,} models) of nearly equal quality may be generated.
This situation derives from the lack of certain critical observables
in the $\chi^{2}$-measure. As we shall see, the particular linear
combination of model parameters defining the soft direction often
provides enough hints to identify the missing observable(s).
Moreover, through this sort of analysis one may establish
correlations among observables that are particularly sensitive
to such soft directions. This is important as certain observables
may be easier to measure than others. A particular topical case
is that of the neutron-skin thickness in ${}^{208}$Pb and the
electric dipole polarizability~\cite{Reinhard:2010wz,
Piekarewicz:2010fa,Tamii:2011pv}.

\subsection{Linear Regression and Covariance Analysis}
\label{LRCA}

As discussed earlier, relativistic models of nuclear structure
are characterized by a number of model parameters, such as
masses, Yukawa couplings, and non-linear meson coupling
constants. Following the notation of Ref.~\cite{Reinhard:2010wz},
we denote a point in such a parameter space by
${\bf p}\!=\!(p_{1},\ldots,p_{F})$, where $F$ is the total
number of model parameters. In principle, each value of ${\bf p}$
represents a model. In practice, of course, one is ordinarily
interested in the {\sl ``best model''} as defined by a quality
measure. To do so, the model parameters are often calibrated to
a well-determined set of ground-state properties of finite nuclei
(such as masses and charge radii) that is supplemented by a few
bulk properties of infinite nuclear matter (such as the binding
energy, incompressibility coefficient, and symmetry energy at
saturation density).
Once the model parameters and the group of observables have
been selected, the optimal parameter set is determined via a
least-squares fit to the following $\chi^{2}$ quality measure:
%%%
\begin{equation}
 \chi^{2}({\bf p}) = \sum_{n=1}^{N}
 \left(\frac{\mathcal{O}_{n}^{\rm (th)}({\bf p})-
 \mathcal{O}_{n}^{\rm (exp)}}
 {\Delta\mathcal{O}_{n}}\right)^{2} \;.
 \label{ChiSquare}
\end{equation}
%%%
Here $N$ (often much larger than $F$) denotes the total number of
selected observables whereas ``th'' and ``exp'' stand for the
theoretical prediction and experimental measurement, respectively.
Further, every observable is weighted by a factor of
$(\Delta\mathcal{O}_{n})^{-1}$ that is (customarily) associated with
the accuracy of the measurement.

We assume that---through a numerical procedure that is not of
particular relevance to this work---an accurately-calibrated model
${\bf p}_{0}$ has been found. This implies that all first derivatives
of $\chi^{2}$ vanish at ${\bf p}_{0}$. That is,
%%%
\begin{equation}
 \frac{\partial\chi^{2}({\bf p})}{\partial p_{i}}
 \Big|_{{\bf p}={\bf p}_{0}} \equiv
 \partial_{i}\chi^{2}({\bf p}_{0})=0 \quad
 ({\rm for}~i=1,\ldots,F)\;.
\end{equation}
%%%
The existence of the minimum (as opposed to a maximum or saddle
point) also implies that a particular set of $F$ second derivatives
(to be defined shortly) must all be positive. Approaches based on a
least-squares fit to a $\chi^{2}$-measure often culminate with the
identification of the optimal parametrization ${\bf p}_{0}$. The
predictive power of the model may then be appraised by computing
observables that were not included in the fitting protocol. Less
often, however, least-squares-fit approaches are used to evaluate
the {\sl ``uniqueness''} of the model. In other words, how fast does
the $\chi^{2}$-measure deteriorate as one moves away from ${\bf
p}_{0}$? Clearly, if the minimum is relatively flat (at least along
one direction), then there will be little (or no) deterioration in
the quality of the fit. Through a statistical analysis, we will be
able to obtain a physically reasonable domain of parameters.  We
implement such analysis by studying the small oscillations around
the $\chi^{2}$-minimum.  As a bonus, we will be able to uncover
correlations among observables and attach meaningful theoretical
error bars to the theoretical predictions~\cite{Reinhard:2010wz}. To
start, we expand the $\chi^{2}$-measure around the optimal
${\bf p}_{0}$ model. That is,
%%%
\begin{equation}
 \chi^2({\bf p}) = \chi^{2}({\bf p}_{0}) + \frac{1}{2}\sum_{i,j=1}^{F}
 ({\bf p}-{\bf p}_{0})_{i} ({\bf p}-{\bf p}_{0})_{j}
 \partial_{i}\partial_{j}\chi^{2}({\bf p}_{0}) + \ldots
 \label{Taylor1}
\end{equation}
%%%
For convenience, we quantify the departure from the minimum
by defining scaled, dimensionless variables
%%%
\begin{equation}
  x_{i} \equiv \frac{({\bf p}-{\bf p}_{0})_{i}}{({\bf p}_{0})_{i}} \;.
 \label{xDef}
\end{equation}
%%%
In terms of these scaled variable, the quadratic deviations of
the $\chi^{2}$-measure from its minimum value take the
following compact form:
%%%
\begin{equation}
 \chi^2({\bf p}) -\chi^{2}({\bf p}_{0})
 \equiv \Delta\chi^2({\bf x}) =
 {\bf x}^{T}{\hat{\mathcal M}}\,{\bf x} \;,
 \label{Taylor2}
\end{equation}
%%%
where ${\bf x}$ is a column vector of dimension $F$,
${\bf x}^{T}$ is the corresponding transpose (row)
vector, and ${\hat{\mathcal M}}$ is the {\sl symmetric}
$F\!\times\!F$ matrix of second derivatives defined by
%%%
\begin{equation}
 {\mathcal M}_{ij} = \frac{1}{2}\left(\frac{\partial\chi^{2}}
 {\partial x_{i}\partial x_{j}}\right)_{{\bf x}=0} =
 \frac{1}{2} ({\bf p}_{0})_{i}({\bf p}_{0})_{j}
 \partial_{i}\partial_{j}\chi^{2}({\bf p}_{0}) \;.
\label{MMatrix}
\end{equation}
%%%
Being symmetric, the matrix ${\hat{\mathcal M}}$ can be brought
to a diagonal form by means of an orthogonal (change-of-basis)
transformation. Denoting by ${\hat{\mathcal A}}$ the orthogonal
matrix whose columns are composed of the normalized eigenvectors
and by ${\hat{\mathcal D}}\!=\!{\rm diag}
(\lambda_{1},\ldots,\lambda_{F})$ the {\sl diagonal} matrix of
eigenvalues, the following relation holds true:
${\hat{\mathcal M}} = {\hat{\mathcal A}}{\hat{\mathcal D}}
 {\hat{\mathcal A}^{T}}$. By inserting this relation into
Eq.~(\ref{Taylor2}), we obtain the following simple and
illuminating expression:
%%%
\begin{equation}
  \Delta\chi^2({\bf x}) =
  {\bf x}^{T}\Big(
   {\hat{\mathcal A}}{\hat{\mathcal D}}{\hat{\mathcal A}^{T}}
   \Big) {\bf x} ={\bm\xi}^{T}
   {\hat{\mathcal D}}{\bm\xi}=\sum_{i=1}^{F}
   \lambda_{i}\xi_{i}^{2} \;.
 \label{Taylor3}
\end{equation}
%%%
Here the vector ${\bm\xi}\!=\!{\hat{\mathcal A}^{T}}{\bf x}$
represents a point in parameter space expressed, not in terms of the
original model parameters ($g_{\rm s},g_{\rm v},\ldots$) but rather,
in terms of the new ({\sl ``rotated''}) basis.  As previously
advertised, the deviations of the $\chi^{2}$-measure from its minimum
value have been parametrized in terms of $F$ uncoupled harmonic
oscillators---with the eigenvalues playing the role of the spring
constants. In this way, each eigenvalue controls the deterioration in
the quality of the fit as one moves along a direction defined by its
corresponding eigenvector.  A {\sl ``soft''}
direction---characterized by a small eigenvalue and thus little
deterioration in the $\chi^{2}$ measure---involves a particular linear
combination of model parameters that is poorly constrained by the
choice of observables included in the least-squares fit.
By isolating such linear combination(s) one can identify what kind
of observables ({\sl e.g.,} isovector observables) should be added to
the $\chi^{2}$-measure to better constrain the theoretical model.
Moreover, one may also explore correlations among various
observables  ({\sl e.g.,} neutron-skin thickness and dipole
polarizability) thereby facilitating the experimental extraction of
 some of these critical observables. This could be done by either
refining existing experimental measurements or by designing brand
new ones.

A concept of fundamental importance to the correlation analysis is
the {\sl covariance} between two observables $A$ and $B$, denoted
by ${\rm cov}(A,B)$~\cite{Brandt:1999}. Assuming that
$({\bf x}^{(1)},\ldots,{\bf x}^{(M)})$ represent $M$ points (or models)
in the neighborhood of the optimal model ${\bf x}^{(0)}\!=\!{\bf 0}$,
the covariance between $A$ and $B$ is defined as:
%%%
\begin{equation}
 {\rm cov}(A,B) = \frac{1}{M} \sum_{m=1}^{M}
 \left[\Big(A^{(m)}-\langle A\rangle\Big)
 \Big(B^{(m)}-\langle B\rangle\Big)\right] =
 \langle AB\rangle\ -
 \langle A\rangle\langle B\rangle\;,
\label{Covariance1}
\end{equation}
%%%
where $A^{(m)}\!\equiv\!A({\bf x}^{(m)})$ and
``$\langle\,\rangle$'' denotes a statistical average. From the above
definition the {\sl correlation coefficient}---often called the
Pearson product-moment correlation coefficient---now follows:
%%%
\begin{equation}
 \rho(A, B) = \frac{{\rm cov}(A,B)}
 {\sqrt{ {\rm var}(A) {\rm var}(B) }} \;,
 \label{Correlation}
\end{equation}
%%%
where the {\sl variance} of $A$ is simply given by ${\rm
var}(A)\!\equiv\!{\rm cov}(A,A)$. Note that two observables are said
to be fully correlated if $\rho(A,B)\!=\!1$, fully anti-correlated
if $\rho(A,B)\!=\!-1$, and uncorrelated if $\rho(A,B)\!=\!0$.  If
one expands the deviation of both observables from their average
value, then {\rm cov}(A, B) may be written as
%%%
\begin{equation}
 {\rm cov}(A,B) = \sum_{i,j=1}^{F}
 \frac{\partial A}{\partial x_{i}}
 \left[\frac{1}{M}\sum_{m=1}^{M}
   x_{i}^{(m)}x_{j}^{(m)}\right]
 \frac{\partial B}{\partial x_{j}}\equiv
 \sum_{i,j=1}^{F}
 \frac{\partial A}{\partial x_{i}}C_{ij}
 \frac{\partial B}{\partial x_{j}}\;,
\label{Covariance2}
\end{equation}
%%%
where both derivatives are evaluated at the minimum
(${\bf x}^{(0)}\!=\!{\bf 0}$) and the {\sl covariance matrix}
$C_{ij}$ has been introduced~\cite{Brandt:1999}.
In order to continue, it is critical to decide how should the
$M$ points be generated.  A particularly convenient choice is to
assume that these $M$ points (or models) are distributed
according to the quality measure $\chi^{2}({\bf x})$. That is,
we assume a probability distribution $\phi({\bf x})$ given by
%%%
\begin{equation}
 \phi({\bf x}) = \exp\left[-\frac{1}{2}\Delta\chi^2({\bf x})\right] =
  \exp\left(-\frac{1}{2} {\bf x}^{T}{\hat{\mathcal M}}\,
  {\bf x}\right) \;.
 \label{Phi}
\end{equation}
%%%
The covariance matrix may then be written as follows:
%%%
\begin{equation}
 C_{ij} = \frac{\int x_{i}x_{j}\phi({\bf x}) d{\bf x}}
 {\int \phi({\bf x}) d{\bf x}} =
 \frac{1}{Z(0)}\left[\frac{\partial^{2}Z({\bf J})}
 {\partial J_{i}\partial J_{j}}\right]_{{\bf J}={\bf 0}} \;,
 \label{Covariance3}
\end{equation}
%%%
where we have defined the {\sl ``partition''} function
$Z({\bf J})$ as
%%%
\begin{equation}
  Z({\bf J}) =
  \int\phi({\bf x}) e^{{\bf J}\cdot{\bf x}} d{\bf x} =
  \int \exp\left(-\frac{1}{2}
  {\bf x}^{T}{\hat{\mathcal M}}\,{\bf x}+
  {\bf J}\cdot{\bf x}\right) d{\bf x}\;.
\label{PartitionFcn1}
\end{equation}
%%%
The above gaussian integrals may be readily evaluated
by completing the square. One obtains
%%%
\begin{equation}
  Z({\bf J}) = Z(0)\exp\left(\frac{1}{2}
  {\bf J}^{T}{\hat{\mathcal M}^{-1}}\,{\bf J}\right)
  \equiv Z(0)e^{W({\bf J})} \;.
\label{PartitionFcn2}
\end{equation}
%%%
Hence, under the assumption that the model parameters are
generated according to the $\chi^{2}$-measure, the covariance
matrix becomes equal to the inverse of the matrix of second
derivatives of $\chi^{2}$. That is,
%%%
\begin{equation}
 C_{ij} =
 \frac{1}{Z(0)}\left[\frac{\partial^{2}Z({\bf J})}
 {\partial J_{i}\partial J_{j}}\right]_{{\bf J}={\bf 0}} =
 \frac{\partial^{2}W({\bf J})}{\partial J_{i}\partial J_{j}} =
 ({\mathcal M}^{-1})_{ij}
\label{Covariance4}
\end{equation}
%%%
Finally then, we arrive at a form for the covariance of two
observables that is both simple and easy to compute:
%%%
\begin{equation}
 {\rm cov}(A,B) = \sum_{i,j=1}^{F}
 \frac{\partial A}{\partial x_{i}}
  (\hat{{\mathcal M}}^{-1})_{ij}
 \frac{\partial B}{\partial x_{j}} =
 \sum_{i=1}^{F}
 \frac{\partial A}{\partial \xi_{i}}
 \lambda_{i}^{-1}
 \frac{\partial B}{\partial \xi_{i}} \;.
 \label{Covariance5}
\end{equation}
%%%
The last term in the previous expression is particularly illuminating.
Consider, for example, the case of a very soft direction in the
$\chi^{2}$-measure, namely, an eigenvector of $\hat{\mathcal M}$
(say ${\bm\xi}_{i}$) with a very small eigenvalue (say
$\lambda_{i}^{-1}\!\gg\!1$).  Such a situation routinely emerges in RMF
models whenever two or more isovector parameters are included in the
Lagrangian density but only masses and charge---not neutron---radii
are used to define the $\chi^{2}$-measure.  Having identified a soft
direction, one could then search for an observable $A$ ({\sl e.g.,}
the neutron-skin thickness in ${}^{208}$Pb) that is particularly
sensitive to such a soft direction (as indicated by ${\partial
A}/{\partial\xi_{i}}\!\gg\!1$).  Adding such an observable to the
$\chi^{2}$-measure will stiffen the formerly soft direction, thereby
improving the predictive power of the model. Moreover, if $A$ is
difficult to measure, one could search for alternative observables
that are strongly correlated to $A$. Although some of these notions
have been heuristically implemented for some time, the statistical
analysis discussed here provides a quantitative measure of the
correlation between observables~\cite{Reinhard:2010wz}.

\section{Results}
\label{Results}
In this section we provide two simple examples that illustrate the
ideas presented in the previous sections.  Here terms such as {\sl
``unique''} and {\sl ``predictive''} will be used to
characterize a model. We regard a model as being unique if {\sl all}
the eigenvalues of $\hat{\mathcal M}$ are large ({\sl i.e.,}
$\lambda_{i}\!\gg\!1$ for all $i$). A model is predictive if it can
successfully account for physical observables not included in the
$\chi^{2}$-measure. Note that a model has been defined here as
consisting of both an underlying Lagrangian density (or effective
interaction) and a set of physical observables defining the
$\chi^{2}$-measure.

\subsection{Example 1: Linear Walecka Model}
\label{Walecka1}
We start this section by discussing the linear Walecka model as
an example of a model that is unique but not predictive. The
Lagrangian density for this case is simple as it only contains two
coupling constants~\cite{Walecka:1974qa,Serot:1984ey}. That is,
%%%
\begin{equation}
{\mathscr L}_{\rm int} = g_{\rm s}\bar\psi\psi\phi
 - g_{\rm v}\bar\psi\gamma^{\mu}\psi V_\mu\;.
\label{LWalecka}
\end{equation}
%%%
The Walecka model is perhaps the simplest model that can
account---at  the mean-field level---for the saturation of
symmetric nuclear matter. Indeed, it is the saturation density
and the energy per nucleon at saturation that are typically
used to calibrate the two parameters of the model.
To make this simple model slightly less trivial we
determine the two parameters of the model by minimizing
a quality measure $\chi^{2}$ defined in terms of three
{\sl ``observables''}: (i) the saturation density $\rho_{0}$,
(ii) the energy per nucleon at saturation $\varepsilon_{0}$,
and (iii) the effective Dirac mass $M_{0}^{\star}$. Central
values and uncertainties for these three quantities are
given as follows:
%%%
\begin{subequations}
 \begin{align}
  & \rho_{0} = (0.155 \pm 0.01)~{\rm fm}^{-3}\;, \\
  & \varepsilon_{0} = (-16 \pm 1)~{\rm MeV} \;, \\
  & M_{0}^{\star} = (0.6 \pm 0.1)~M \;.
 \end{align}
\label{Database}
\end{subequations}
%%%
Using standard numerical techniques, a minimum value for
the $\chi^2$-measure of $\chi_{0}^{2}\!=\!0.34145$ is obtained
at
%%%
\begin{subequations}
\begin{align}
   g_{\rm s}^{2} = \phantom{1}93.62647 \;, \\
   g_{\rm v}^{2} = 180.48347 \;.
  \label{WaleckaMin}
\end{align}
\end{subequations}
%%%
Having computed the minimum value of the quality measure, we now
examine its behavior around the minimum. This
is implemented by diagonalizing the symmetric matrix of second
derivatives $\hat{\mathcal M}$ [see Eq.~(\ref{MMatrix})].  The outcome
of such a diagonalization procedure is the diagonal matrix of
eigenvalues $\hat{\mathcal D}$ and the orthogonal matrix of normalized
eigenvectors $\hat{\mathcal A}$. That is,
%%%
\begin{subequations}
\begin{align}
  & \hat{\mathcal D}={\rm diag}(\lambda_{1},\lambda_{2})=
      {\rm diag}(7.4399\!\times\!10^{4},8.3195\!\times\!10^{1}) \;,
      \label{DMatrix}\\
  & \hat{\mathcal A}=
      \left(\begin{array}{rr}
      \cos\theta & \sin\theta  \\
    -\sin\theta & \cos\theta
      \end{array}\right) =
      \left(\begin{array}{rr}
      0.74691 & 0.66492  \\
    -0.66492 & 0.74691
     \end{array}\right) \;.
      \label{AMatrix}
\end{align}
\end{subequations}
%%%
It is evident that both eigenvalues are very large. This indicates
that both directions in parameter space are stiff and consequently
the quality measure ($\Delta\chi^{2}
\!=\!\lambda_{1}\xi_{1}^{2}+\!\lambda_{2}\xi_{2}^{2}$) will
deteriorate rapidly as one moves away from the
$\chi^{2}$-minimum. Note that $\lambda_{1}$ is significantly larger
than $\lambda_{2}$; this is to be expected.  When probing the
parameter landscape along the first direction (i.e., $\xi_{2}\!=\!0$)
the scalar and vector coupling constants move {\sl ``out-of-phase''}
(see the first column of the matrix $\hat{\mathcal A}$). For example,
the scalar attraction would get larger at the same time that the
vector repulsion would get smaller. This would yield a significant
increase in the binding energy per particle and consequently a
drastic deterioration in the $\chi^{2}$-measure.  Recall that large
and {\sl cancelling} scalar and vector potentials are the hallmark of
relativistic mean-field models.

To quantify the extent by which the linear Walecka model is unique,
we now proceed to compute the variance in the coupling constants
using Eq.~(\ref{Covariance5}). We obtain
%%%
\begin{subequations}
 \begin{align}
 & \sigma_{\rm s}^{2} = \left(\hat{\mathcal M}^{-1}\right)_{11} =
     \lambda_{1}^{-1}\cos^{2}\theta+\lambda_{2}^{-1}\sin^{2}\theta =
     5.3217\!\times\!10^{-3} \;, \\
 & \sigma_{\rm v}^{2} = \left(\hat{\mathcal M}^{-1}\right)_{22} =
     \lambda_{1}^{-1}\sin^{2}\theta+\lambda_{2}^{-1}\cos^{2}\theta =
     6.7116\!\times\!10^{-3} \;.
 \end{align}
 \label{VarianceSV2}
\end{subequations}
%%%
In turn, this translates into the following uncertainties in the optimal
values of the coupling constants:
%%%
\begin{subequations}
 \begin{align}
   & g_{\rm s}^{2} = \phantom{1}93.62647\,(1\pm\sigma_{\rm s})
   = \phantom{1}93.62647\pm 6.83008 \;, \\
   & g_{\rm v}^{2} = 180.48347\,(1\pm\sigma_{\rm v})
   = 180.48347\pm 14.78596 \;.
 \end{align}
 \label{VarianceSV3}
\end{subequations}
%%%
We conclude that the uncertainties in the model parameters---and thus
in most of the predictions of the model---are of the order of 5-to-10
percent. In principle, the model uncertainties could be reduced by
refining the experimental database [see Eq.\,(\ref{Database})].  The
great merit of the present statistical approach is that one may
systematically explore the extent by which the experimental
measurement must be refined in order to achieve the desired
theoretical accuracy. Note that the theoretical uncertainties are
dominated by the {\sl smallest} eigenvalue of $\hat{\mathcal M}$ [see
Eq.\,(\ref{VarianceSV2})].  Thus, assessing the uniqueness of the
model by varying each model-parameter individually ({\sl e.g,} first
$g_{\rm s}^{2}$ and then $g_{\rm v}^{2}$) is misleading and ill
advised.  It is misleading because in doing so the quality measure
will in general be dominated by the {\sl largest} eigenvalue [see
Eq.\,(\ref{Taylor3})].  Yet it is the lowest eigenvalue that
determines the uniqueness of the model.

\begin{figure}[ht]
\vspace{-0.05in}
\includegraphics[width=0.75\columnwidth,angle=0]{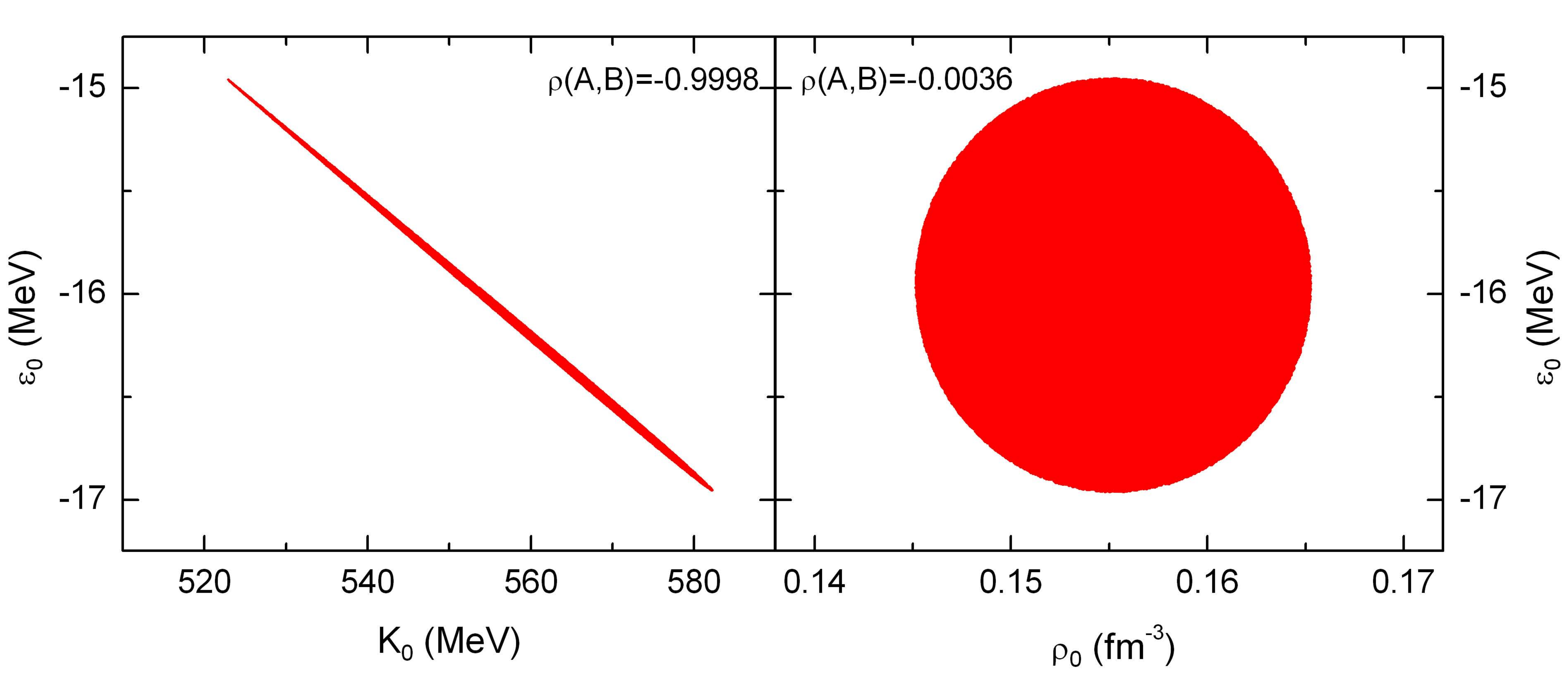}
\caption{(Color online) Predictions from the linear Walecka models
 for the saturation density, binding energy, and incompressibility
 coefficient at saturation. Model parameters were generated according
 to the distribution $\exp(-\Delta\chi^{2}/2)$. Both of the covariance
 ellipsoids were generated by limiting the models to the region
 $\Delta\chi^{2}\!\le\!1$.}
\label{Fig1}
\end{figure}

Carrying out the covariance analysis further, we now proceed to
compute correlation coefficients between model parameters
and observables [see Eqs.\,(\ref{Correlation}) and~(\ref{Covariance5})].
In estimating uncertainties in the model parameters one concentrates
on the diagonal elements of the (inverse) matrix of second derivatives
[see Eq.\,(\ref{VarianceSV2})]. Information on the correlation between
model parameters is, however, stored in the off-diagonal elements.
For example, the correlation coefficient between $g_{\rm s}^{2}$ and
$g_{\rm v}^{2}$ is given by
%%%
\begin{equation}
  \rho(g_{\rm s}^{2}, g_{\rm v}^{2}) =
  \frac{\left(\hat{\mathcal M}^{-1}\right)_{12}}
  {\sqrt{\left(\hat{\mathcal M}^{-1}\right)_{11}
           \left(\hat{\mathcal M}^{-1}\right)_{22}}}= 0.9977 \;.
 \label{CorrSV}
\end{equation}
%%%
The strong (positive) correlation between $g_{\rm s}^{2}$ and
$g_{\rm v}^{2}$ is easily understood.  Given that configurations in
parameter space are distributed according to the $\chi^{2}$-measure,
model parameters in which $g_{\rm s}^{2}$ and $g_{\rm v}^{2}$ move
out-of-phase are strongly suppressed, as they are controlled by
the largest eigenvalue $\lambda_{1}$. As a result, an overwhelming
number of configurations are  generated with $g_{\rm s}^{2}$ and
$g_{\rm v}^{2}$ moving in phase, thereby leading to a large
positive correlation.  Correlation coefficients between various
isoscalar observables have been tabulated in Table~\ref{Table1}.
Given that the correlation coefficients are sensitive to the first
derivatives of the observables along all (eigen)directions [see
Eq.~(\ref{Covariance5})], we have listed them for completeness in
Table~\ref{Table2}. We observe that all observables display a much
larger sensitivity to the stiff direction than to the soft one. This
could (and does) lead to sensitive cancellations since the large
derivatives compensate for the small value of
$\lambda_{1}^{-1}$. Indeed, the correlation between the saturation
density and the binding energy at saturation is very small. On the
other hand, the incompressibility coefficient appears to be strongly
correlated to the binding energy. This behavior is also displayed in
graphical form in Fig.~\ref{Fig1} where predictions for the various
observables were generated with model parameters distributed
according to $\phi({\bf x})$ [see Eq.~(\ref{Phi})]. Note that the
{\sl covariance ellipsoids} in Fig.~\ref{Fig1} were generated by
selecting those model parameters that satisfy
$\Delta\chi^{2}\!\le\!1$.
%%%%%%%%%%%%%%%%%%%%%%%%%%%%%%%%%%%%%%%%%%%%%%%%%%%%%%%%%%%%%%%%%
  \begin{table}[h]
  \begin{tabular}{|c|c|c|c|c|c|}
    \hline
      & $\varepsilon_0$  & $\rho_0$   & $K_{0}$ & $M^{\star}_{0}$   \\
    \hline
    $\varepsilon_{0}$     & $+1.0000$ & $-0.0036$  & $-0.9998$  & $+0.8867$ \\
    \hline
    $\rho_{0}$                & $-0.0036$ & $+1.0000$  & $-0.0174$  & $+0.4591$  \\
    \hline
    $K_{0}$                     & $-0.9998$  & $-0.0174$  & $+1.0000$  & $+0.8962$  \\
    \hline
    $M^{\star}_{0}$           & $+0.8867$ & $+0.4591$  & $+0.8962$  & $+1.0000$  \\
    \hline
  \end{tabular}
 \caption{Correlation coefficients between isoscalar observables
   in the linear Walecka model.}
 \label{Table1}
 \end{table}
%%%%%%%%%%%%%%%%%%%%%%%%%%%%%%%%%%%%%%%%%%%%%%%%%%%%%%%%%%%%%%%%%

%%%%%%%%%%%%%%%%%%%%%%%%%%%%%%%%%%%%%%%%%%%%%%%%%%%%%
%%%%%%%%%%%%%%%%%%%%%%%%%%%%%%%%%%%%%%%%%%%%%%%%%%%%%

%%%%%%%%%%%%%%%%%%%%%%%%%%%%%%%%%%%%%%%%%%%%%%%%%%%%%
\begin{widetext}
 \begin{table}[h]
  \begin{tabular}{|c||l|l|}
   \hline
    & $\hfill\partial\xi_{1}\hfill$ & $\hfill\partial\xi_{2}\hfill$ \\
    \hline
    \hline
    $\partial\varepsilon_{0}$  & $-1.6698\!\times\!10^{1}$
                                            & $-1.1716\!\times\!10^{-1}$   \\
    \hline
    $\partial\rho_{0}$ & $\phantom{+}3.6619$
                                & $-5.7333\!\times\!10^{-1}$   \\
    \hline
    $\partial K_{0}$    & $\phantom{+}1.4261\!\times\!10^{1}$
                               & $\phantom{+}1.1055\!\times\!10^{-1}$   \\
    \hline
    $\partial M^{\ast}_{0}$   & $-3.2349$  & $-8.8817\!\times\!10^{-2}$   \\
    \hline
  \end{tabular}
 \caption{First derivatives of the {\sl scaled} observables
  ({\sl i.e.,} observable scaled to its value at the
  $\chi^{2}$-minimum) as a function of $\xi_{1}$ and
  $\xi_{2}$ evaluated at the $\chi^{2}$-minimum;  see
  Eq.~(\ref{Covariance5}).}
 \label{Table2}
 \end{table}
 \end{widetext}
%%%%%%%%%%%%%%%%%%%%%%%%%%%%%%%%%%%%%%%%%%%%%%%%%%%%%
Based on the previous statistical analysis it appears that the linear
Walecka model is unique (at least at the $5\!-\!10$\% level).  But is
the linear Walecka model predictive?  To test the predictability of
the model we focus on two physical observables that were not
included in the $\chi^{2}$-measure, namely, the incompressibility
coefficient $K_{0}$ and the symmetry energy $J$. We obtain---with
properly computed theoretical errors---the following results:
%%%
\begin{subequations}
 \begin{align}
    K_{0} & = (552.537 \pm 29.655)~{\rm MeV}\;,  \\
         J  & = (19.775 \pm 0.683)~{\rm MeV} \;.
 \end{align}
 \label{LinearKJ}
\end{subequations}
%%%
Both predictions, even after theoretical errors have been
incorporated, differ significantly from the presently acceptable
values of $K_{0}\!\approx\!(240\pm20)\,$MeV and
$J\!\approx\!(32\pm2)\,$MeV. This conclusion should hardly come as a
surprise. After all, the predominant role played by the model
parameters $\kappa$ and $\lambda$ in softening the incompressibility
coefficient and $g_{\rho}$ in stiffening the symmetry energy have been
known for a long time.  What is relevant from the present statistical
analysis is that we have established quantitatively that the linear
Walecka model fails because its prediction for $K_{0}$ differs from
the experimental value by more than 10 standard deviations.
We must then conclude that whereas the linear Walecka is (fairly)
unique, it is not predictive. We now proceed to discuss a particular
extension of the Walecka model that is highly predictive but not
unique: the non-linear FSUGold model.

\subsection{Example 2: Non-linear FSUGold Model}
\label{FSUGold}

Modern relativistic models of nuclear structure have evolved
significantly since the early days of the linear Walecka model.
In the present example we focus on the FSUGold parameter
set~\cite{Todd-Rutel:2005fa} that is defined by an interacting
Lagrangian density of the following form:
%%%
\begin{equation}
 {\mathscr L}_{\rm int} = \bar\psi \left[g_{\rm s}\phi
    \!-\! \left(g_{\rm v}V_\mu
    \!+\! \frac{g_{\rho}}{2}\mbox{\bm{$\tau$}}\cdot{\bf b}_{\mu}
    \!+\! \frac{e}{2}(1\!+\!\tau_{3})A_{\mu}\right)\gamma^{\mu}
    \right]\psi - \frac{\kappa}{3!} \Phi^3 \!-\!\frac{\lambda}{4!}{\Phi}^4
     \!+\!\frac{\zeta}{4!}\Big(W_{\mu}W^\mu\Big)^2
     \!+\!\Lambda_{\rm v}\Big(W_{\nu}W^\nu\Big)
     \Big({\bf B}_{\mu}\cdot{\bf B}^{\mu}\Big) \;.
\label{FSULagrangian}
\end{equation}
%%%
Modifications to the linear Walecka model are motivated by the
availability of an ever increasing database of high-quality data.  For
example, the two non-linear scalar terms $\kappa$ and $ \lambda$
induce a significant softening of the compression modulus of nuclear
matter relative to the original Walecka
model~\cite{Walecka:1974qa,Boguta:1977xi,Serot:1984ey}. This is
demanded by measurement of the giant monopole resonance in medium to
heavy nuclei~\cite{Youngblood:1999}. Further, omega-meson
self-interactions, as described by the parameter $\zeta$, also serve
to soften the equation of state of symmetric nuclear matter but at
much higher densities. Indeed, by tuning the value of $\zeta$ it is
possible to produce maximum neutron star masses that differ by almost
one solar mass while maintaining the saturation properties of nuclear
matter intact~\cite{Mueller:1996pm}.  Such a softening appears
consistent with the dynamics of high-density matter as probed by
energetic heavy-ion collisions~\cite{Danielewicz:2002pu}.  Finally,
$\Lambda_{\rm v}$ induces isoscalar-isovector mixing and is
responsible for modifying the poorly-constrained density dependence of
the symmetry energy~\cite{Horowitz:2000xj,Horowitz:2001ya}.  In
particular, a softening of the symmetry energy induced by
$\Lambda_{\rm v}$ appears consistent with the distribution of both
isoscalar monopole and isovector dipole strength in medium to heavy
nuclei~\cite{Piekarewicz:2000nm,
Piekarewicz:2003br,Todd-Rutel:2005fa}. In summary, FSUGold is a fairly
successful RMF model that has been validated against theoretical,
experimental, and observational constraints~\cite{Piekarewicz:2007dx}.
Note that as additional laboratory and observational data become
available (notably the recent report of a 2-solar mass neutron
star~\cite{Demorest:2010bx}) refinements to the model my be
required~\cite{Fattoyev:2010mx}. For now, however, we will be
content with using the FSUGold model to study the small oscillations
around the minimum.

As mentioned earlier, a model should be understood as a combination of
an interacting Lagrangian density and a quality measure. We define the
$\chi^{2}$-measure in terms of the following set of observables
generated directly from the FSUGold parameter set:
%%%
\begin{subequations}
 \begin{align}
  & \rho_{0} = 0.1484~{\rm fm}^{-3}\;, \\
  & \varepsilon_{0} = -16.30~{\rm MeV} \;, \\
  & \varepsilon(2\rho_{0}) = -5.887~{\rm MeV} \;, \\
  & K_{0} = 230.0~{\rm MeV} \;, \\
  & M_{0}^{\star} =  0.6100~M \;, \\
  & \tilde{J} = 26.00~{\rm MeV} \;, \\
  & L = 60.52~{\rm MeV} \;, \\
  & M_{\rm max} = 1.722~M_{\odot} \;.
 \end{align}
\label{Database2}
\end{subequations}
%%%
Note that in all cases a 2\% uncertainty is attached to all
observables---except in the case of the slope of the symmetry energy
$L$ where the significant larger value of 20\% is assumed.  This
reflects our poor understanding of the density dependence of the
symmetry energy. Also note that $\tilde{J}$ represents the value of
the symmetry energy at a sub-saturation density of
$\rho\!\approx\!0.1\,{\rm fm}^{-3}$---a density at which the
theoretical uncertainties are minimized~\cite{Furnstahl:2001un}.
Finally, notwithstanding the Demorest {\sl et al.}
result~\cite{Demorest:2010bx}, the maximum neutron star mass is fixed
at $M_{\rm max} \!=\!1.722~M_{\odot}$. Given that a theoretical model
is used to generate the various observables, a much larger database
could be used to define the $\chi^{2}$-measure, if desired.  By
construction, a very small value for the $\chi^{2}$-measure is
obtained at the FSUGold minimum. We now proceed to explore the wealth
of information available as one studies deviations around this minimum
value.  As in the previous section, the symmetric matrix of second
derivatives $\hat{\mathcal M}$ (now a $7\!\times\!7$ matrix) may be
diagonalized by means of an orthogonal transformation. The diagonal
matrix of eigenvalues $\hat{\mathcal D}$ and the matrix of
eigenvectors $\hat{\mathcal A}$ are given by
%%%
\begin{footnotesize}
\begin{subequations}
\begin{align}
  & \hat{\mathcal D}={\rm diag}
      (1.2826\!\times\!10^{6},
       1.5305\!\times\!10^{4},
       4.2472\!\times\!10^{2},
       3.2113\!\times\!10^{2},
       1.2692\!\times\!10^{2},
       6.9619, 3.7690)\;,
     \label{DMatrix2}\\
  & \hat{\mathcal A}=
      \left(\begin{array}{rrrrrrr}
      -7.4967\!\times\!10^{-1} & -2.3685\!\times\!10^{-1} &
        3.0853\!\times\!10^{-1} &  1.2931\!\times\!10^{-1}  &
      -5.1254\!\times\!10^{-1} & -8.5089\!\times\!10^{-2} &
        6.8417\!\times\!10^{-3}\\
        6.5682\!\times\!10^{-1} & -1.5751\!\times\!10^{-1} &
        3.6654\!\times\!10^{-1} &  1.6504\!\times\!10^{-1} &
      -6.0281\!\times\!10^{-1} & -1.3685\!\times\!10^{-1} &
        8.5353\!\times\!10^{-3}\\
      -1.5331\!\times\!10^{-4} &  3.1315\!\times\!10^{-3} &
      -7.0050\!\times\!10^{-1} &  6.8701\!\times\!10^{-2} &
      -3.9206\!\times\!10^{-1} & -3.3843\!\times\!10^{-2} &
        5.9137\!\times\!10^{-1}\\
       3.8535\!\times\!10^{-2} & -2.8770\!\times\!10^{-1} &
     -2.4254\!\times\!10^{-2} &  4.7416\!\times\!10^{-1} &
       4.3796\!\times\!10^{-2} &  8.2968\!\times\!10^{-1} &
     -5.7643\!\times\!10^{-3}\\
       3.9417\!\times\!10^{-2} & -6.8525\!\times\!10^{-1} &
     -1.3772\!\times\!10^{-1} &  3.8776\!\times\!10^{-1} &
       3.5428\!\times\!10^{-1} & -4.8376\!\times\!10^{-1} &
       2.6431\!\times\!10^{-3}\\
      -5.9458\!\times\!10^{-2} &  6.0558\!\times\!10^{-1} &
        5.4897\!\times\!10^{-2} &  7.5689\!\times\!10^{-1} &
        6.8021\!\times\!10^{-2} & -2.2175\!\times\!10^{-1} &
        6.2795\!\times\!10^{-3}\\
       1.2995\!\times\!10^{-4} & -3.1465\!\times\!10^{-3} &
       5.0714\!\times\!10^{-1} & -5.7010\!\times\!10^{-2} &
       2.9691\!\times\!10^{-1} &  3.6238\!\times\!10^{-2} &
       8.0628\!\times\!10^{-1}
    \end{array}\right) \;.
    \label{AMatrix2}
\end{align}
\end{subequations}
\end{footnotesize}
%%%
Note that the scaled parameters of the model are associated to the
original coupling constants as follows:
%%%
\begin{equation}
  \{x_1, x_2, x_3, x_4, x_5, x_6, x_7\} \rightarrow
  \{g_{\rm s}^2, g_{\rm v}^2,g_{\rho}^2, \kappa, \lambda,
  \zeta, \Lambda_{\rm v}\} \;.
\end{equation}
%%%
We observe that the stiffest direction is dominated by two
isoscalar parameters and represents---as in the case of the
linear Walecka model---an out-of-phase oscillation between
the scalar attraction and the vector repulsion. Given that in
RMF models the cancellation between the scalar attraction and
the vector repulsion is so delicate, any out-of-phase motion
yields a significant change in the binding energy per nucleon
and a correspondingly dramatic increase in the quality measure.
The second stiffest direction also involves exclusively isoscalar
parameters and is dominated by the quartic scalar ($\lambda$)
and vector ($\zeta$) couplings---and to a lesser extent by the
cubic term ($\kappa$). This linear combination of parameters
is largely constrained by the incompressibility coefficient
$K_{0}$ and the maximum neutron-star mass $M_{\rm max}$.
Although the determination of the maximum neutron-star mass
to a 2\% accuracy presents a significant observational challenge,
our statistical analysis suggests that such a determination
would strongly constrain the equation of state from saturation density
up to neutron-star densities. The third stiffest direction (with still
a fairly large eigenvalue of $\lambda_{3}\!\approx\!425$) is dominated
by the two isovector parameters $g_{\rho}^{2}$ and $\Lambda_{\rm
v}$. For this particular {\sl ``mode''} both parameters oscillate out
of phase. This behavior can be readily understood by recalling the
expression for the symmetry energy~\cite{Horowitz:2001ya}:
%%%
\begin{equation}
  E_{\rm sym}(\rho) = \frac{k^{2}_{F}}{6 E_F^{\ast}} +
  \frac{g_{\rho}^{2}\rho}{8m_{\rho}^{\ast 2}}\;, \quad
  \Big(m_{\rho}^{\ast 2} \equiv m_{\rho}^{2}+
   2\Lambda_{\rm v}g_{\rho}^{2}W_{0}^{2}\Big) \;.
\end{equation}
%%%
In order for the symmetry energy $\tilde{J}$ to remain fixed, then both
$g_{\rho}^{2}$ and $\Lambda_{\rm v}$ must move in phase. If they move
out of phase, then the symmetry energy can not be kept at this value
and the quality measure deteriorates. By the same
token, the in-phase motion of $g_{\rho}^{2}$ and $\Lambda_{\rm v}$ is
very poorly constrained---as evinced by the last and softest
direction.  And it is only because the slope of the symmetry energy
$L$ was assumed to be somehow constrained (at the 20\% level) that a
positive eigenvalue was even obtained. Note that one of the main goals
of the successfully commissioned Lead Radius experiment (PREx) at the
Jefferson Laboratory is to constrain the density dependence of the
symmetry energy ({\sl i.e.,} $L$) by accurately measuring the neutron
radius of $^{208}$Pb~\cite{Horowitz:1999fk,Michaels:2005}.
The next to last eigenvalue ($\lambda_{6}\!\approx\!7$) is
also relatively small. This suggest that the out-of-phase motion of
the two non-linear scalar couplings ($\kappa$ and $\lambda$) is
poorly constrained by the nuclear-matter observables defining the
quality measure. Perhaps supplementing the quality measure with
finite-nuclei observables will help ameliorate this problem. Work
along these lines is currently in progress.

We now proceed to estimate theoretical uncertainties as well as
to compute correlation coefficients for both the model parameters
and the physical observables. We start by computing theoretical
uncertainties ({\sl i.e.,} variances) for the model parameters. These
are given by [see Eq.\,(\ref{Covariance5})]
%%%
\begin{equation}
  \sigma_{i}^{2} = \Big(\hat{\mathcal M}^{-1}\Big)_{ii} =
  \Big(\hat{\mathcal A}\hat{\mathcal D}^{-1}
          \hat{\mathcal A}^{T}\Big)_{ii} =
   \sum_{j=1}^{7} {\mathcal A}^{2}_{ij}\lambda_{j}^{-1} \;,
 \label{Sigmas}
\end{equation}
%%%
and result in the following theoretical uncertainties for the model
parameters:
%%%
\begin{subequations}
 \begin{align}
    g_{\rm s}^{2} &= 112.19955\pm 6.54468 \;\; [5.833\%] \;,\\
    g_{\rm v}^{2} &= 204.54694\pm 15.81183 \;\; [7.730\%] \;,\\
    g_{\rho}^{2}  &= 138.47011\pm 42.75427 \;\; [30.876\%] \;,\\
   \kappa & = 1.42033\pm 0.44827\;\; [31.561\%]\;,\\
   \lambda & = 0.02376\pm 0.00445\;\; [18.748\%]\;,\\
   \zeta & =      0.06000\pm 0.0057 \;\; [9.447\%]\;,\\
   \Lambda_{\rm v} & = 0.03000\pm 0.01251 \;\; [41.711\%]\;.
 \end{align}
 \label{VarianceFSU1}
\end{subequations}
%%%
We observe that three out of the five isoscalar parameters, namely,
$g_{\rm s}^{2}$, $g_{\rm v}^{2}$, and $\zeta$, are relatively well
constrained (at the $\lesssim\!10$\% level).  Whereas $g_{\rm s}^{2}$
and $g_{\rm v}^{2}$ are well determined by the saturation properties
of symmetric nuclear matter, it is the maximum neutron-star mass that
constrains $\zeta$. Yet the remaining two isoscalar parameters
($\kappa$ and $\lambda$) are poorly determined. This is particularly
true in the case of $\kappa$ which displays a large ($\approx\!30$\%)
uncertainty.  As alluded earlier, these large uncertainties develop
because the out-of-phase motion of $\kappa$ and $\lambda$---as
controlled by the relatively soft sixth eigenvector---is poorly
constrained.  Given that the in-phase motion of the two isovector
parameters ($g_{\rho}^{2}$ and $\Lambda_{\rm v}$) is controlled by the
softest of eigenvectors, the theoretical uncertainties in these
parameters is also fairly large ($\approx\!30$\% and $\approx\!40$\%,
respectively). However, whereas the reason for the latter is
associated with the large error bars assigned to $L$, we are unaware
at this time on how to better constrain $\kappa$ and $\lambda$.
Perhaps supplementing the quality measure with information on
various nuclear compressional modes may help resolve this issue.
Plans to do so in the near future are under consideration.

%%%%%%%%%%%%%%%%%%%%%%%%%%%%%%%%%%%%%%%%%%%%%%%%%%%%%
\begin{figure}[ht]
\vspace{-0.05in}
\includegraphics[width=0.6\columnwidth,angle=0]{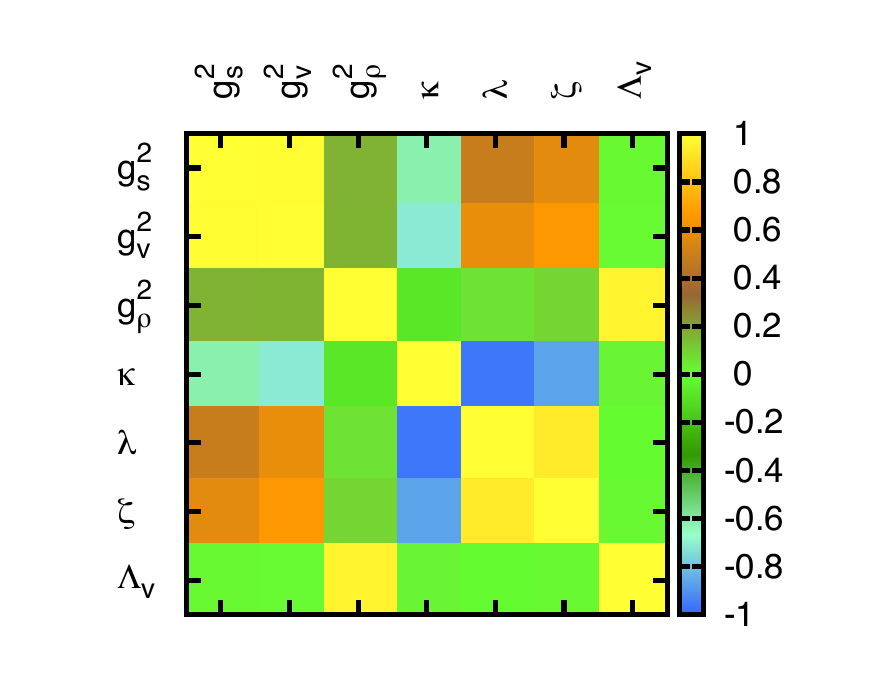}
\caption{(Color online) Color-coded plot of the 21 independent
 correlation coefficients between the 7 model parameters of the
 FSUGold effective interaction.}
\label{Fig2}
\end{figure}
%%%%%%%%%%%%%%%%%%%%%%%%%%%%%%%%%%%%%%%%%%%%%%%%%%%%%

We have computed correlation coefficients between all 21 distinct
pairs of model parameters and have displayed them in graphical
(color-coded) form in Fig.~\ref{Fig2}. As depicted in the figure, the
strongest correlations are between $g_{\rm s}^{2}$ and
$g_{\rm v}^{2}$~(0.988), $g_{\rho}^{2}$ and $\Lambda_{\rm v}$~(0.967), and
$\kappa$ and $\lambda$~(-0.962).  As alluded in the case of the
simpler linear Walecka model, the correlations are dominated by
the {\sl softest directions}; in this case the sixth and seventh
eigenvectors.  Given that for these two eigenvectors
$g_{\rm s}^{2}$ and $g_{\rm v}^{2}$ as well as $g_{\rho}^{2}$ and
$\Lambda_{\rm v}$ move in phase whereas $\kappa$ and $\lambda$
move out of phase, the observed correlations ensue. In other words,
the three largest eigenvalues strongly suppress the generation of
model parameters with $g_{\rm s}^{2}$ and $g_{\rm v}^{2}$ moving out
of phase, $\kappa$ and $\lambda$ in phase, and $g_{\rho}^{2}$ and
$\Lambda_{\rm v}$ out of phase, respectively.  Note then, that the
distribution of isovector parameters $g_{\rho}^{2}$ and $\Lambda_{\rm
v}$ is generated in such a way that the symmetry energy at
sub-saturation density $\tilde{J}$ remains fixed at 26 MeV (at least
within a 2\% uncertainty). This quantitative fact validates our
heuristic approach---already employed numerous times---to
correlate isovector observables (see Refs.~\cite{Piekarewicz:2007us,
Piekarewicz:2009gb,Piekarewicz:2010fa} and references therein).

We now extend the covariance analysis to the case of physical
observables. To do so, we must supply the relevant {\sl ``matrix''}
of first derivatives [see Eq.\,(\ref{Covariance5})].  For completeness
we list the first derivatives in tabular form in
Table~\ref{Table3}.  Note that in the case of the model parameters
the corresponding matrix of first derivatives is the matrix of
eigenvectors $\hat{\mathcal A}$.  The derivatives encapsulate the
sensitivity of the various observables to changes along the different
eigenvectors. For example, whereas isoscalar observables (such as
$\varepsilon_{0}, \rho_{0}, K_{0}$) are insensitive to changes
along the mostly isovector seventh eigenvector, both $L$ and the
neutron-skin thickness of ${}^{208}$Pb, $R_{n}\!-\!R_{p}$, display a
fairly large sensitivity.
%%%%%%%%%%%%%%%%%%%%%%%%%%%%%%%%%%%%%%%%%%%%%%%%%%%%%%%%%%%%%%%%%
  \begin{widetext}
  \begin{table}[h]
  \begin{tabular}{|c||l|l|l|l|l|l|l|}
    \hline
    & $\hfill\partial\xi_1\hfill$  & $\hfill\partial\xi_2\hfill$
    & $\hfill\partial\xi_3\hfill$  & $\hfill\partial\xi_4\hfill$
    & $\hfill\partial\xi_5\hfill$ & $\hfill\partial\xi_6\hfill$
    & $\hfill\partial\xi_7\hfill$  \\
    \hline
    \hline
    $\partial\varepsilon_0$ &
    $\phantom{+}1.0551\!\times\!10^{+1}$ &
    $-7.1882\!\times\!10^{-1}$ &
    $-2.9433\!\times\!10^{-2}$ &
    $\phantom{+}4.9029\!\times\!10^{-2}$ &
    $\phantom{+}5.1626\!\times\!10^{-2}$ &
    $-1.7025\!\times\!10^{-2}$ &
    $-7.7009\!\times\!10^{-4}$ \\
    \hline
    $\partial\rho_0$ & $-2.2472$  &
    $\phantom{+}1.3904$  &
    $-9.8868\!\times\!10^{-4}$ &
    $\phantom{+}2.2220\!\times\!10^{-1}$ &
    $\phantom{+}5.3440\!\times\!10^{-2}$ &
    $\phantom{+}1.8998\!\times\!10^{-2}$ &
    $\phantom{+}1.5875\!\times\!10^{-3}$   \\
    \hline
    $\partial K_0$ & $-7.4792$  &
    $\phantom{+}1.3890$  &
    $-3.6799\!\times\!10^{-2}$ &
    $-1.8808\!\times\!10^{-1}$ &
    $\phantom{+}4.8215\!\times\!10^{-2}$ &
    $-2.1691\!\times\!10^{-2}$ &
    $-3.0813\!\times\!10^{-4}$   \\
    \hline
    $\partial M^{\ast}$ & $\phantom{+}1.1505$  &
    $-5.5581\!\times\!10^{-1}$ &
    $-1.0614\!\times\!10^{-1}$ &
    $-8.3170\!\times\!10^{-2}$ &
    $\phantom{+}1.6492\!\times\!10^{-1}$ &
    $\phantom{+}1.2614\!\times\!10^{-2}$ &
    $-2.7701\!\times\!10^{-3}$   \\
    \hline
    $\partial\tilde{J}$   &
    $-2.7862\!\times\!10^{-1}$ &
    $\phantom{+}6.5586\!\times\!10^{-2}$ &
    $-3.9136\!\times\!10^{-1}$ &
    $\phantom{+}2.5698\!\times\!10^{-2}$ &
    $-6.8299\!\times\!10^{-2}$ &
    $\phantom{+}6.0862\!\times\!10^{-4}$ &
    $\phantom{+}6.2323\!\times\!10^{-4}$   \\
    \hline
    $\partial J$ & $-1.6811$  &
    $\phantom{+}9.4897\!\times\!10^{-1}$ &
    $-3.2446\!\times\!10^{-1}$ &
    $\phantom{+}1.7826\!\times\!10^{-1}$ &
    $-3.6792\!\times\!10^{-2}$ &
    $\phantom{+}6.2151\!\times\!10^{-3}$ &
    $-8.4827\!\times\!10^{-2}$   \\
    \hline
    $\partial L$ & $-1.8759$  &
    $\phantom{+}1.1812$  &
    $\phantom{+}1.5849\!\times\!10^{-2}$ &
    $\phantom{+}2.6611\!\times\!10^{-1}$ &
    $-1.0180\!\times\!10^{-1}$ &
    $-2.2526\!\times\!10^{-2}$ &
    $-3.8593\!\times\!10^{-1}$   \\
    \hline
    $\partial(R_n\!-\!R_p)$ &
    $\phantom{+}7.0224$  &
    $\phantom{+}9.6188\!\times\!10^{-2}$ &
    $-2.3362\!\times\!10^{-1}$ &
    $\phantom{+}1.9804\!\times\!10^{-1}$ &
    $-2.5651\!\times\!10^{-2}$ &
    $-1.9368\!\times\!10^{-2}$ &
    $-3.4167\!\times\!10^{-1}$   \\
    \hline
    $\partial R_{1.0}$ &
    $\phantom{+}9.9947\!\times\!10^{-1}$ &
    $-3.1989\!\times\!10^{-1}$ &
    $-1.6534\!\times\!10^{-2}$ &
    $-6.5939\!\times\!10^{-2}$ &
    $-4.5197\!\times\!10^{-2}$ &
    $-4.5964\!\times\!10^{-3}$ &
    $-3.9800\!\times\!10^{-2}$   \\
    \hline
    $\partial R_{1.4}$ &
    $\phantom{+}5.0300\!\times\!10^{-1}$ &
    $-3.0884\!\times\!10^{-1}$ &
    $\phantom{+}9.3778\!\times\!10^{-3}$ &
    $-1.1119\!\times\!10^{-1}$ &
    $-6.3792\!\times\!10^{-2}$ &
    $\phantom{+}3.1016\!\times\!10^{-3}$ &
    $-2.6571\!\times\!10^{-2}$   \\
    \hline
    $\partial M_{\rm max}$ &
    $-2.7675\!\times\!10^{-1}$ &
    $-1.4882\!\times\!10^{-1}$ &
    $\phantom{+}3.1173\!\times\!10^{-2}$ &
    $-1.6394\!\times\!10^{-1}$ &
    $-6.8790\!\times\!10^{-2}$ &
    $\phantom{+}3.4817\!\times\!10^{-2}$ &
    $-2.4367\!\times\!10^{-3}$   \\
    \hline
  \end{tabular}
  \caption{First derivatives of the {\sl scaled} observables
  ({\sl i.e.,} observable scaled to its value at the
  $\chi^{2}$-minimum) as a function of $\xi_{i}$ at the
  $\chi^{2}$-minimum;  see Eq.~(\ref{Covariance5}).}
 \label{Table3}
 \end{table}
 \end{widetext}
%%%%%%%%%%%%%%%%%%%%%%%%%%%%%%%%%%%%%%%%%%%%%%%%%%%%%%%%%%%%%%%%%

Given the enormous interest in constraining the density dependence of
the symmetry energy, we estimate theoretical uncertainties on
three---mostly isovector---observables.  These are the symmetry energy
at saturation density $J$, the neutron-skin thickness of ${}^{208}$Pb,
and the radius ($R_{1.4}$) of a $M\!=\!1.4 M_{\odot}$ neutron
star. Recall that it was $\tilde{J}$ (not $J$) that was included in the
definition of the quality measure. We obtain,
 %%%
\begin{subequations}
 \begin{align}
   &  J  = (32.593 \pm 1.574)\,{\rm MeV} \;\; [4.830\%] \;,\\
   & R_{n}\!-\!R_{p}  = (0.207\pm 0.037)\,{\rm fm} \;\;  [17.698\%]\;, \\
  & R_{1.4} =(11.890 \pm 0.194)\,{\rm km} \;\;  [1.631\%] \;.
 \end{align}
 \label{VarianceFSU2}
\end{subequations}
%%%
%%%%%%%%%%%%%%%%%%%%%%%%%%%%%%%%%%%%%%%%%%%%%%%%%%%%%
\begin{figure}[ht]
\vspace{-0.05in}
\includegraphics[width=0.6\columnwidth,angle=0]{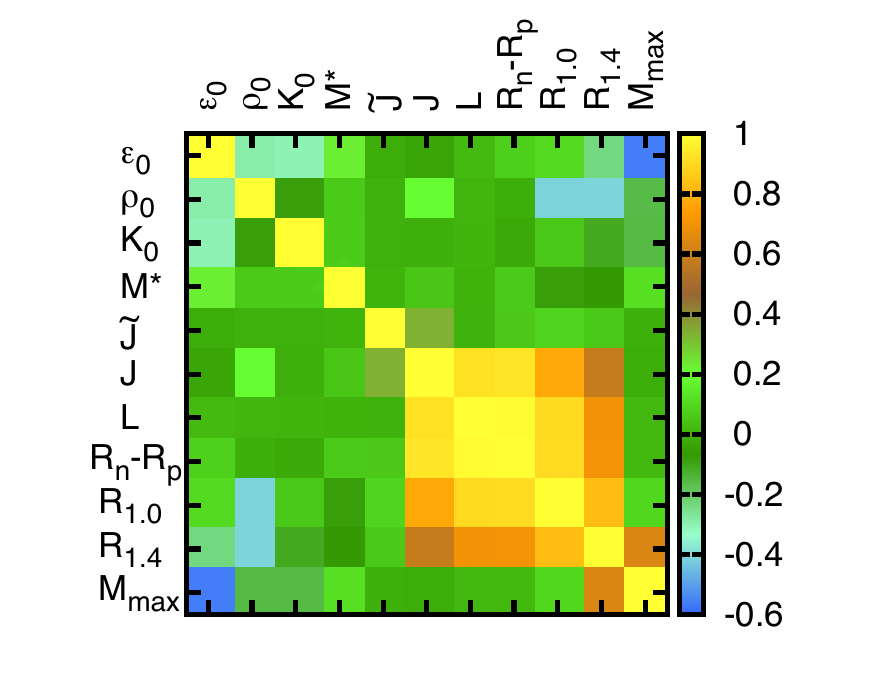}
\caption{(Color online) Color-coded plot of the 55 independent
 correlation coefficients between 11 physical observables as
 computed with the FSUGold effective interaction.}
\label{Fig3}
\end{figure}
%%%%%%%%%%%%%%%%%%%%%%%%%%%%%%%%%%%%%%%%%%%%%%%%%%%%%
We now comment on each of these cases individually. Before we do so,
however, note that correlation coefficients for 11 observables ({\sl
i.e.,} 55 independent pairs) are depicted in a color-coded format in
Fig.~\ref{Fig3}. First, the central value of $J$ along with its
theoretical uncertainty may be easily understood by invoking a
first-order expansion for the symmetry energy $\tilde{J}$ at
sub-saturation density ($\tilde{\rho}_{0}\!\approx\!0.103\,{\rm
fm}^{-3}$) in terms of $J$ and $L$~\cite{Piekarewicz:2008nh}. That is,
%%%
\begin{equation}
  J = \tilde{J} + xL + \ldots
 \approx (32.208 \pm 1.346) \,{\rm MeV}\;, \quad
  x = \frac{1}{3}\left(1-\frac{\tilde{\rho}_{0}}{\rho_{0}}\right)
                   \approx 0.103 \;,
 \label{Japprox}
\end{equation}
%%%
where the errors were added in quadrature.
So although $J$ is strongly correlated to $L$ (with a correlation
coefficient of 0.922) the error in the former is significantly smaller
than the latter because of the small value of $x$.  Second, for the
neutron-skin thickness of ${}^{208}$Pb we find a theoretical error
comparable to the one assumed for $L$ and a correlation coefficient
between the two observables of almost one (0.995).  Such a strong
correlation is consistent with two recent studies that employ a large
number of accurately-calibrated relativistic and non-relativistic
interactions to uncover the
correlation~\cite{Centelles:2008vu,RocaMaza:2011pm}.  Also consistent
with these studies, specifically with Ref.~\cite{RocaMaza:2011pm}, is
the fact that the proposed 1\% measurement of the neutron radius of
${}^{208}$Pb by the PREx collaboration~\cite{Horowitz:1999fk,
Michaels:2005} may not be able to place a significant constrain on
$L$. For example, our covariance analysis suggests that the 20\%
uncertainty assumed for $L$ translates into a theoretical error in the
neutron skin of $0.037\,{\rm fm}$---or about a 0.7\% uncertainty in
the neutron radius of ${}^{208}$Pb.  Conversely, if $L$ is to be
determined to within 10\% ({\sl i.e.,} $L\!\approx\!60\pm 6\,{\rm
MeV}$) then the neutron skin must be constrained to about $0.018\,{\rm
fm}$ so the neutron radius must be measured with close to a 0.3\%
accuracy---a fairly daunting task.  Finally, we obtain a very small
theoretical uncertainty for the radius of a 1.4 solar-mass neutron
star and a correlation coefficient between $L$ and $R_{1.4}$ (or
$R_{n}\!-\!R_{p}$ and $R_{1.4}$) of 0.811.  Although the
radius of the neutron star is sensitive to the density dependence of
the symmetry energy~\cite{Steiner:2004fi}, $R_{1.4}$ can not be
uniquely constrained by a measurement of $R_{n}\!-\!R_{p}$ because
whereas the latter depends on the symmetry energy at (or below)
saturation density, the former is also sensitive to the equation of
state at higher densities~\cite{Horowitz:2001ya}. Note that a far
better correlation coefficient of 0.942 is obtained between $L$ and
the radius of a 1.0 solar-mass neutron star. Regardless (with all
things being equal) knowledge of the slope of the symmetry energy
to a 20\% accuracy significantly constrains the stellar radius.

\section{Conclusions}
\label{Conclusions}

The demand for theoretical predictions that include meaningful
and reliable uncertainties is increasing. Such a sentiment has been
articulated in a recent publication by the editors of the Physical
Review A~\cite{PhysRevA.83.040001}. The need to quantify model
uncertainties in an area such as theoretical nuclear physics is
particularly urgent as models that are fitted to experimental data are
then used to extrapolate to the extremes of temperature, density,
isospin asymmetry, and angular momentum.
Inspired by some of the central ideas developed in
Ref.~\cite{Reinhard:2010wz}, a systematic statistical approach was
applied to a class of relativistic mean-field models.  The aim of this
statistical analysis was twofold. First, to attach meaningful and
reliable theoretical uncertainties to both the model parameters as
well as to the predicted observables. Second, to quantify the degree
of correlation between physical observables.

Modern relativistic mean-field models have evolved considerably since
the early days of the linear Walecka model.  Based on certain
shortcoming of the Walecka model---most notably the inability to
reproduce the incompressibility coefficient of symmetric nuclear
matter---the Lagrangian density was augmented by non-linear cubic and
quartic scalar-meson terms. However, based on modern
effective-field-theory tenets, such as naturalness and power counting,
a consistent Lagrangian density should include all terms up to fourth
order in the meson fields. But in doing so, how should one constrain
the large number of model parameters?  In principle, one should follow
the standard protocol of determining all model parameters through a
$\chi^{2}$-minimization procedure.  In practice, however, many
successful theoretical approaches arbitrarily set some of the model
parameters to zero.  The argument behind this fairly ad-hoc procedure
is that the full set of parameters is poorly determined by existing
data, so ignoring a subset model parameters does not compromise the
quality of the fit.

A covariance analysis such as the one implemented here should be able
to clarify in a quantitative fashion the precise meaning of a {\sl
``poorly determined set of parameters''}.  To do so, one should
focus---not on the minimum of the $\chi^{2}$-measure but
rather---on its behavior around the minimum. As in any small-oscillations
problem, the deviations around the minimum are controlled by a
symmetric matrix of second derivatives that may be used to extract
theoretical error bars and to compute correlation coefficients among
physical observables. However, to access the wealth of information
available in the covariance analysis we took it a step further and
diagonalized the matrix of second derivatives.  Upon diagonalization,
the deviations of the $\chi^{2}$-measure from the minimum are
parametrized in terms of a collection of {\sl ``uncoupled harmonic
oscillators''}.  By doing so, one could readily identify stiff and
soft modes in parameter space, namely, eigenvectors characterized by
either large or small eigenvalues, respectively.

We now summarize some of the most important lessons learned.
First, a stiff direction represents a particular linear combination of model
parameters that is well constrained by the set of physical observables
included in the $\chi^{2}$-measure. By the same token, a soft
direction suggests that additional physical observables are required
to further constrain the model. Second, given that model parameters
around the minimum are distributed according to the
$\chi^{2}$-measure, the soft directions dominate the correlation
analysis. Finally, testing whether a model is well constrained by
individually varying its parameters---rather than by varying them
coherently as suggested by the structure of the eigenvectors---may be
misleading.  To illustrate these findings we used two relatively
simple, yet illuminating, examples: (a) the linear Walecka model and
(b) the FSUGold parametrization.  Note that ultimately we aim to
implement the covariance analysis with a $\chi^{2}$-measure defined by
a consistent Lagrangian density.

A particularly clear example of a stiff direction was represented by
the out-of-phase motion of the scalar $g_{\rm s}$ and vector $g_{\rm
v}$ coupling constants in the linear Walecka model. Indeed, increasing
the scalar attraction while at the same time reducing the vector
repulsion leads to a significant increase in the binding energy per
nucleon and, thus, in a significant deterioration of the
$\chi^{2}$-measure.  The in-phase motion of $g_{\rm s}$ and
$g_{\rm v}$, however, is not as well constrained (the ratio of the two
eigenvalues is about 1000). Therefore, configurations in
parameter space generated by the $\chi^{2}$-measure were dominated
by pairs of coupling constants that were in phase, thereby resulting in
a correlation coefficient between $g_{\rm s}$ and $g_{\rm v}$ that was,
as expected, large and positive. Note, however, that if $g_{\rm s}$
and $g_{\rm v}$ were varied individually, one would erroneously
conclude that the model is much better constrained than it really
is---since changes in the $\chi^{2}$-measure would be
dominated by the largest eigenvalue.

In our second example we considered the accurately-calibrated FSUGold
interaction with an isovector interaction determined by two parameters
($g_{\rho}$ and $\Lambda_{\rm v}$). We found the out-of-phase motion
of $g_{\rho}$ and $\Lambda_{\rm v}$ to be strongly constrained by the
value of the symmetry energy at a density of about $0.1$~fm. However,
our poor knowledge of the density dependence of the symmetry energy
left the in-phase motion of $g_{\rm v}$ and $\Lambda_{\rm v}$ largely
unconstrained. Effectively then, correlations in the isovector sector
were induced by the in-phase motion of $g_{\rho}$ and
$\Lambda_{\rm v}$---subject to the constraint that the symmetry energy
at $\rho\!\approx\!0.1$~fm remains intact. This procedure validates
the heuristic approach that we have used for some time to estimate
correlations among isovector observables.  Yet a benefit of the
present analysis is that one can precisely quantify the theoretical
errors as well as the correlation among observables. For example,
we concluded that if the slope of the symmetry energy is to be
determined with a 10\% uncertainty, then the neutron-skin thickness
of ${}^{208}$Pb should be measured with a 0.3\% accuracy. This more
stringent limit seems to agree with the conclusions of
Ref.~\cite{RocaMaza:2011pm}.

In the future we aim to apply the covariance analysis discussed here
to the construction of a relativistic density functional that will
include all terms up to fourth order in the meson fields. Moreover, we
plan to calibrate the $\chi^{2}$-measure using various properties of
finite nuclei and neutron stars. In addition, we reiterate a point made
in Ref.~\cite{Reinhard:2010wz} that the methodology used in this work
should be applicable to any problem where model parameters are
determined from optimizing a quality measure.

\begin{acknowledgments}
 We thank Profs. W. Nazarewicz and P.-G. Reinhard for many useful
 conversations.  We also thank Prof. M. Riley for calling our attention
 to Ref.~\cite{PhysRevA.83.040001}. This work was supported in part
 by the United States Department of Energy under grant
 DE-FG05-92ER40750.
\end{acknowledgments}

\bibliography{../ReferencesJP}

\end{document}